\documentclass[10pt,sigconf]{acmart}
\graphicspath{ {./images/}}
\usepackage{physics}
\usepackage{float}
\usepackage{comment}
\usepackage{xcolor}
\setcopyright{none}
\usepackage{threeparttable}
\usepackage{multirow}

\usepackage[linesnumbered,vlined,ruled,commentsnumbered]{algorithm2e}

\AtBeginDocument{%
  }

\acmYear{2026}

\renewcommand\footnotetextcopyrightpermission[1]{}

\title{Flow-Based Lattice Surgery Optimization with Runtime T Gate Scheduling}

\begin{document}
\raggedbottom

\author{Raymond Iacobacci}
\email{ra586960@ucf.edu}
\orcid{0009-0000-8783-2704}
\affiliation{%
  \institution{University of Central Florida}
  \city{Orlando}
  \state{Florida}
  \country{USA}
}

\author{Tianyi Hao}
\email{tianyi.hao@wisc.edu}
\orcid{????}
\affiliation{%
  \institution{University of Wisconsin - Madison}
  \city{Madison}
  \country{USA}}

\author{Neer Patel}
\email{ne004475@ucf.edu}
\orcid{????}
\affiliation{%
  \institution{University of Central Florida}
  \city{Orlando}
  \state{Florida}
  \country{USA}
}

\author{Siyuan Niu}
\email{siyuan.niu@ucf.edu}
\orcid{????}
\affiliation{%
  \institution{University of Central Florida}
  \city{Orlando}
  \state{Florida}
  \country{USA}
}

\typeout{COLUMNWIDTH=\the\dimexpr\columnwidth}

\begin{abstract}

Lattice surgery on the surface code is a leading route to fault-tolerant quantum computation. Compiling logical circuits into lattice surgery operations is a pivotal step that determines the resource cost of a computation. Prior compilers build on idealized assumptions, and one that works in practical settings remains to be developed. Existing works are either unscalable or route one gate at a time within a layer, so the embedding depends on the order of gates and opportunities to route many gates at once are lost. In addition, they assume magic states are always available and cost-free, even though preparing one costs more than a Clifford operation. 

We present \emph{FlowRouter}, the first topological lattice surgery compiler that places Clifford routing and stochastic magic state cultivation inside a single 3D embedding. At compile time, FlowRouter formulates the routing of an entire circuit layer as a maximum-flow problem, embedding many gates simultaneously. At runtime, it allows magic states to be cultivated at every idling patch, works with the realistic cultivation process with multiplexing and multi-stage post-selections, and absorbs the stochasticity by introducing delay mechanisms that add as little spacetime volume as possible. With magic states assumed free, FlowRouter reduces spacetime volume by 2.3$\times$ and compiles 10.8$\times$ faster than the state-of-the-art static topological lattice surgery compiler. With the cultivation-aware setting that includes both Clifford and magic state cost, FlowRouter reduces spacetime volume by 2.3$\times$ on a distance-13 surface code when compared with the state-of-the-art runtime compiler.

\end{abstract}

\maketitle
\pagestyle{plain}

\section{Introduction}

Quantum computing promises advantages across scientific domains such as quantum chemistry and optimization. However, the ubiquitous noise in today's quantum hardware hinders the realization of practical applications. Quantum error correction (QEC) is widely believed to be the path toward reliable computation: by encoding logical information redundantly, it suppresses physical noise exponentially and thereby enables the fault-tolerant execution needed to unlock quantum advantage.

Among the proposed codes, surface code is one of the most popular choices due to its high threshold and decodability~\cite{fowler2012surface}. Recently, numerous surface-code experiments have been demonstrated across diverse quantum platforms~\cite{google2023suppressing,google2024belowthreshold,bluvstein2024logical}. To execute logical operations fault-tolerantly, lattice surgery is a widely adopted technique that functions by \emph{merging} and \emph{splitting} adjacent code patches~\cite{litinski2019game, horsman2012surface, fowler2019low}. This approach is especially well suited for solid state hardware with limited qubit connectivity, including superconducting and spin qubit systems.

Compiling a logical circuit into lattice surgery introduces substantial overhead in both qubit count and execution time, which the compiler attempts to minimize. This overhead is captured by the \emph{spacetime volume}, of which two measures are in common use: the \emph{bounding volume}, the product of the footprint that the computation reserves and the number of code cycles it runs for, and the \emph{active volume}, which counts only the patches actually in use. The gap between the two measures can be exploited for magic state preparation when magic states are prepared via cultivation rather than via distillation.

Early lattice surgery compilers focus on mapping and routing circuits on a 2D lattice~\cite{molavi2025dascot, beverland2022edpc, leblond2024realistic, hua2021autobraid}, while recent compilers exploit the topological nature of the problem by expressing the circuit as a ZX diagram and mapping it to a 3D pipe diagram~\cite{zhou2026topols,tan2024lassynth,liao2026kovalq}, whose added time dimension offers more routing freedom and thus more compressed spacetime volumes. Three limitations remain. First, every state-of-the-art compiler that scales to a reasonable number of qubits routes gates one at a time~\cite{molavi2025dascot, watkins2024high, leblond2024realistic, beverland2022edpc, silva2024lssp, zhou2026topols}, so its output depends on the order in which gates are considered; the compilers that do commit routes do so via exhaustive search and stop scaling at a few tens of qubits~\cite{tan2024lassynth}. Second, compilers assume magic states are prepared outside the perimeter of the program and are always available, an assumption inherited from distillation factories that magic state cultivation has made obsolete. Third, and as a consequence of the second, no compiler treats Clifford routing and magic state preparation as a single allocation problem, even though under cultivation the two contend for the same idle patches.

\begin{figure}[t]
    \centering
    \includegraphics[width=0.98\columnwidth]{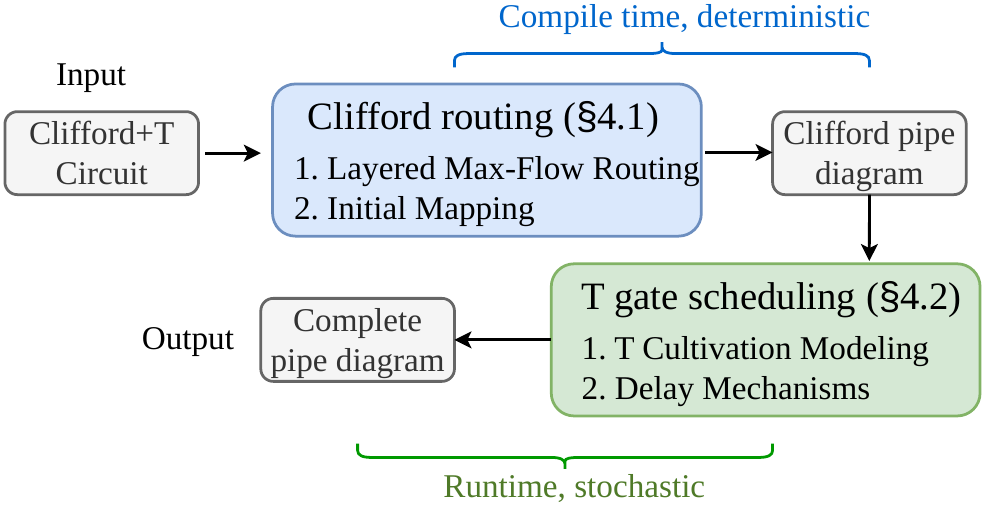}
    \caption{Overview of our proposed lattice surgery compiler FlowRouter.}
    \label{fig:overview}
\end{figure}

To address these limitations, we propose \emph{FlowRouter}, the first complete compiler built on topological pipe diagrams that jointly handles all Clifford gates together with magic state cultivation (MSC)~\cite{gidney2024cultivation} and injection. The overview of FlowRouter is shown in Figure~\ref{fig:overview}. First, FlowRouter solves the routing of all Clifford gates at compile time: it converts the input circuit into a ZX diagram, optimizes the diagram with ZX rewrite rules, and formulates the embedding from the ZX diagram to a 3D topological pipe diagram as a maximum-flow problem. This generates an embedding for multiple gates in one layer simultaneously and produces highly compressed pipe diagrams while scaling to a large number of qubits. Second, FlowRouter allows performing MSC everywhere if possible at runtime and supports the cultivation process with multiplexing and multi-stage post-selections. Since one MSC takes longer than a Clifford operation and is non-deterministic in runtime, the compile time volume may not be able to accommodate it; we therefore introduce multiple delay mechanisms that absorb the MSC latency without inflating the spacetime volume much. We evaluate FlowRouter on a diverse set of benchmarks spanning different quantum algorithms and circuit sizes under two settings: a free-magic setting, in which magic states are assumed instantly available at the program's edge, and a cultivation-aware setting, in which magic states are cultivated in idle code patches. In the free-magic setting, FlowRouter reduces spacetime volume by 2.3$\times$ and compiles $10.8\times$ faster than the state-of-the-art static topological lattice surgery compiler~\cite{zhou2026topols}. In the cultivation-aware setting, it reduces spacetime volume by 2.3$\times$ on a distance-13 surface code over the state-of-the-art runtime scheduler that models cultivation~\cite{hofmeyr2025puremagic}.

Our main contributions are summarized as follows:
\begin{itemize}
\item We present FlowRouter, a lattice surgery compiler based on topological pipe diagrams that, to the best of our knowledge, is the first to place Clifford routing and magic state cultivation inside the same 3D embedding.

\item We formulate Clifford gate routing as a maximum-flow problem which can route multiple gates in the same layer simultaneously and produce highly compressed volumes while remaining fast and scalable to a large number of qubits.

\item We introduce low-latency runtime scheduling for stochastic MSC, using multiple delay mechanisms to integrate magic state preparation and injection into the 3D pipe diagram with minimal additional spacetime volume.

\item We evaluate FlowRouter on a variety of benchmarks, where it achieves substantial spacetime volume reductions in both the free-magic setting and the full-cost setting that charges Clifford routing and stochastic MSC together.

\end{itemize}

\section{Background}
\label{sec:background}

\textbf{Surface code.}
A quantum error-correcting code encodes logical qubits redundantly into physical qubits; its \emph{code distance} $d$ is the minimum number of physical faults that produce an undetectable logical error. The \emph{rotated surface code} that we target encodes one logical qubit into a $d \times d$ array of data qubits interleaved with $d^2 - 1$ measure qubits on a two-dimensional grid with nearest-neighbor interactions only, a natural fit for solid-state platforms such as superconducting and spin qubits. Error correction proceeds in \emph{code cycles}, in each of which every measure qubit extracts an $X$- or $Z$-type stabilizer syndrome; below a threshold physical error rate of roughly $1\%$ the logical error rate is suppressed exponentially in $d$~\cite{fowler2012surface, google2024belowthreshold}. A logical qubit occupies a rectangular \emph{patch} whose two opposite $X$-type and two opposite $Z$-type boundaries are the interface through which every logical operation is performed. Following the $\mathrm{RGB} = XYZ$ convention of the surface code literature and of \texttt{TQEC}~\cite{tqec}, we color $X$ red, $Y$ green, and $Z$ blue throughout.

\textbf{Lattice surgery.}
Lattice surgery~\cite{horsman2012surface, fowler2019low} performs logical operations by deforming patches rather than by transversal gates. Its atomic instruction is the \emph{merge-split}: merging two patches along a shared boundary promotes the idle qubits between them to measure qubits for $d$ code cycles, and splitting them apart again costs one cycle. A merge-split across $Z$ boundaries measures the joint operator $X \otimes X$ and one across $X$ boundaries measures $Z \otimes Z$, so logical qubits that are not adjacent must be linked through intermediate \emph{ancilla} patches. A layout accordingly designates some grid sites as \emph{data patches}, holding the logical qubits of the circuit, and leaves the rest as routing ancillas; a \emph{sparse layout} places data patches on alternating sites so that each keeps free neighbors~\cite{litinski2019game}.

\begin{figure}[t]
    \centering
    \includegraphics[width=0.65\columnwidth]{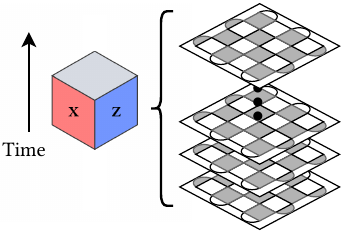}
    \caption{A surface code patch and the cube representing it in a pipe
    diagram, with boundary types carried onto the cube faces.}
    \label{fig:bg-cube}
\end{figure}

\textbf{Pipe diagrams.}

A lattice surgery computation is written as a three-dimensional \emph{pipe diagram}~\cite{gidney2019flexible, tan2024lassynth, tqec}, whose horizontal axes are the qubit grid and whose vertical axis is time. A \emph{cube} is a patch held for $d$ code cycles, its six faces carrying the types of the corresponding boundaries and its two temporal faces giving the bases of initialization and measurement (Figure~\ref{fig:bg-cube}); a \emph{pipe} joining two adjacent cubes is a merge-split, and may carry a Hadamard transition that exchanges the basis of the logical operator crossing it. All cost is borne by the cubes, each occupying $\approx d^3$ spacetime volume: the \emph{bounding volume} of a diagram is the product of its spatial footprint and its temporal depth, and the \emph{active volume} counts only the occupied cubes. Not every arrangement of cubes and pipes is a valid computation; among other local conditions, a pipe must agree in basis with the faces it meets, and the pipes incident on a cube may not span all three axes at once~\cite{tan2024lassynth}.

\begin{figure}[t]
    \centering
    \includegraphics[width=0.98\columnwidth]{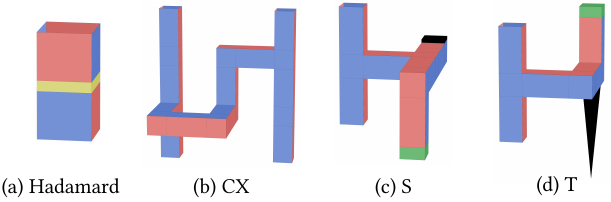}
    \caption{Pipe diagram realizations of the $\{H, S, \mathrm{CX}, T\}$ gate set.}
    \label{fig:gates}
\end{figure}

\textbf{Logical operations.}
We compile circuits over the gate set $\{H, S, \mathrm{CX}, T\}$, realized as shown in Figure~\ref{fig:gates}. Pauli gates are never executed on-chip, but commuted through the circuit and tracked classically at no spacetime cost. $H$ exchanges the two boundary types of a patch. $S$ teleports in an ancilla prepared by in-place $Y$-basis initialization; this initialization comes at a cost of $d \times d \times d/2$ spacetime volume~\cite{gidney2024inplace}. A CX is two merge-splits through a $\ket{+}$ ancilla, measuring $Z \otimes Z$ from the control and then $X \otimes X$ into the target. A $T$ gate teleports in a magic state $\ket{T} \propto \ket{0} + e^{i\pi/4}\ket{1}$, which must be prepared using a specific magic-state preparation protocol such as cultivation or distillation; the gadget leaves behind an $S$ correction conditioned on the injection's measurement outcome, which must therefore be scheduled strictly after the injection it corrects.

\textbf{ZX calculus.}
A ZX diagram~\cite{vandewetering2020zx} is a graph of \emph{spiders}, each of $Z$ (blue) or $X$ (red) type and carrying a phase, joined by wires. A Clifford$+T$ circuit translates into one directly: a CX is a 0-phase $Z$ spider on the control joined to a 0-phase $X$ spider on the target, and $S$ and $T$ are $Z$ spiders of phase $\pi/2$ and $\pi/4$. Pipe diagrams are ZX diagrams in disguise~\cite{debeaudrap2020zx, gidney2019flexible}: every cube is a spider and every pipe a wire, and the color that draws a junction gives its spider type. Compilation is therefore a geometric problem, in which the ZX diagram fixes the graph of spiders that must be realized and the compiler assigns each spider a location in spacetime subject to the validity conditions above. That an embedding realizes the intended computation can be certified by \emph{correlation surfaces}~\cite{tan2024lassynth, debeaudrap2020zx}: sheets threading the pipes of a diagram that relate the logical operators at its inputs to those at its outputs.

\textbf{Magic state cultivation.}
Magic states have traditionally been produced by \emph{distillation}~\cite{bravyi2005universal}, which post-selects many noisy copies to yield one of higher fidelity, at a footprint that confines it to dedicated factories outside the computation~\cite{litinski2019game}. Magic state cultivation (MSC)~\cite{gidney2024cultivation} instead prepares a $T$ state in place within a single logical patch, in three stages. \emph{Injection} prepares an encoded $T$ state at fault distance one. \emph{Cultivation} raises the fault distance to $d_1$ in a small color code through rounds of post-selected syndrome extraction, discarding and restarting the attempt whenever a \emph{postselector} detects an error. \emph{Escape} then grows the result into a distance-$d_2$ surface code patch; post-selection is unavailable after escape, so acceptance is instead decided by the \emph{complementary gap}, a measure of decoder confidence, with attempts below a threshold $G_T$ discarded and restarted. MSC thus trades spacetime volume against output fidelity through $d_1$, $d_2$, and $G_T$, and its completion time is stochastic: whether an attempt yields a usable state is known only at runtime.

\section{Motivation}
\label{sec:motivation}

\subsection{Gate-by-gate Compilation is Suboptimal}
\label{sec:motiv-sequential}

Every lattice surgery compiler that scales to real circuits routes one gate at a time, committing each route along a shortest available path before the next gate is considered~\cite{molavi2025dascot, watkins2024high, leblond2024realistic, beverland2022edpc, silva2024lssp, zhou2026topols}. Sequential compilation makes the combinatorial joint optimization tractable, but it also leaves compilation quality sensitive to the order.

Compilers therefore optimize it based on ordering heuristics~\cite{hua2021autobraid}, simulated annealing~\cite{molavi2025dascot}, and random restarts per layer~\cite{zhou2026topols}, but all of them search a space factorial in the layer width and none removes the dependence on the order.

\begin{figure}[t]
    \centering
    \includegraphics[width=0.98\columnwidth]{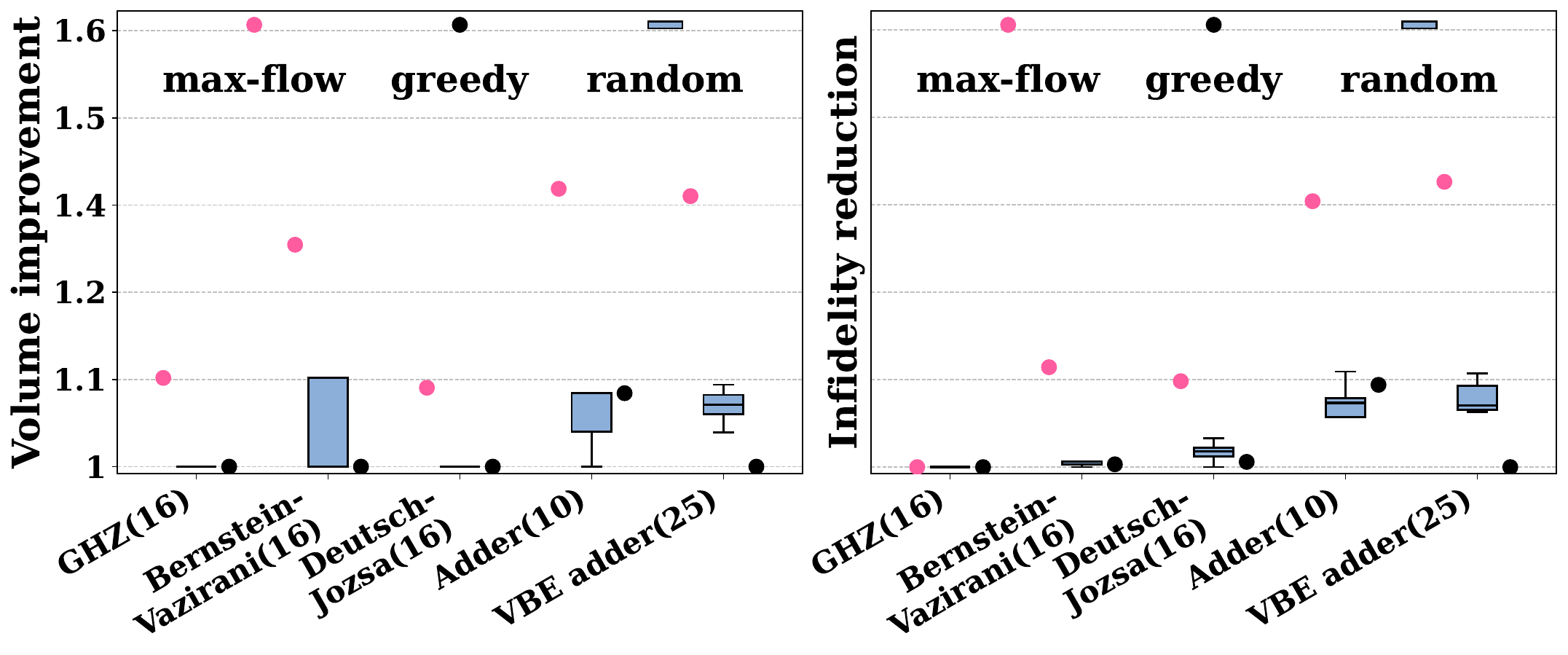}
    \caption{Strict (bounding box around active cubes) spacetime volume and infidelity at $d = 11$ for the same circuits compiled under three embedding strategies: random, greedy, and our max-flow kernel. Metric definitions are given in Section~\ref{sec:eva}. Bernstein-Vazirani(16) and Deutsch-Jozsa(16) are nearly identical to the suite's respective versions and are sourced from TopoLS; they were chosen to build a square layout. There was no substitute for Adder(10).}
    \label{fig:motiv-ordering}
\end{figure}

Figure~\ref{fig:motiv-ordering} measures what that sensitivity costs. We partition each benchmark into layers according to its gate dependencies and embed the gates of every layer in three ways: (1) our kernel embedding based on a max-flow algorithm, which embeds all gates of a layer simultaneously; (2) a greedy embedding, which embeds gates sequentially in a layer in the order of their estimated path distances; and (3) a random embedding, which embeds gates sequentially in a random order. The random embedding is repeated five times while the other two are run once. Every other parameter stays the same.

Compared with the better-performing sequential baseline for each benchmark, our maximum-flow method reduces spacetime volume by 23\% on average and reduces the circuit infidelity by 16\%. These results demonstrate the benefits of routing gates jointly at the layer level instead of committing routes one gate at a time. 

Classical physical VLSI design meets the same wall: nets (wires connecting pins, analogous to gates connecting qubits) routed late inherit the blockages left by nets routed early, with no way to ask an earlier net to keep a region clear~\cite{hu2001survey, albrecht2001global}. The field eventually settled on a change of formulation, routing all nets concurrently as a flow problem over the whole region~\cite{shragowitz1987global, albrecht2001global, kahng2011vlsi}.

\subsection{A layer of CXs is a Maximum Flow}
\label{sec:motiv-maxflow}

Routing gates concurrently is hard in general. Connecting \emph{fixed} terminal pairs by disjoint paths is a multi-commodity problem: minimizing depth is NP-hard~\cite{beverland2022edpc}, and the maximum node-disjoint path set on a grid's subgraph resists approximation to within $2^{\Omega(\sqrt{\log N})}$ under standard assumptions~\cite{chuzhoy2018inapprox}.

A special case is when a data qubit needs a magic state. Since any boundary patch containing a magic state can serve, the destination need not be fixed in advance. A super-source over the data qubits and a super-sink over the boundary turn the instance into a single-commodity unit-capacity maximum flow, solvable exactly in polynomial time~\cite{beverland2022edpc}. Nonetheless, whether this construction can be extended to general CXs was left open~\cite{beverland2022edpc}.

Our observation is that by leveraging the time dimension in the compilation, the CX endpoints are not fixed either. A ZX diagram records which spiders must be connected: a CX is realized by any valid pipe joining a blue junction on the control column to a red junction on the target column. The operations within one layer commute by construction, so the compiler can decide which columns act as sources and which as sinks, and the intractable multi-commodity instance becomes single-commodity. A maximum flow within the layer's spacetime window returns the largest set of mutually non-overlapping valid pipes in a single solve. The freedom leverages the flexibility of 3D lattice surgery compilation; in a 2D time slice, source and sink stay pinned to the two given data patches.

\paragraph{Why not Pauli-based computation.} A well-established alternative commutes the Cliffords to the end and compiles the remaining Pauli-product rotations~\cite{litinski2019game, hofmeyr2025puremagic, silva2024lssp}. It is 2D by design and admits no such formulation: a weight-$k$ rotation must reach $k$ data patches through one ancilla region, reducing to a Steiner tree problem rather than a flow problem~\cite{silva2024lssp}. Clifford$+T$ keeps every multi-qubit operation two-local, so a layer is a set of source--sink pairs and maximum flow applies directly.

\subsection{Magic States Need Runtime Scheduling}
\label{sec:motiv-cultivation}

Prior compilers inherit their treatment of magic states from distillation: a factory is far too large to sit beside the data, so magic states are produced in dedicated regions outside the computation and routed in~\cite{litinski2019game}. Two assumptions come with that placement. First, magic states enter through the perimeter, so a layout of $N$ patches can serve only $\mathcal{O}(\sqrt{N})$ $T$ gates per round of delivery~\cite{beverland2022edpc}. Second, a factory can be replicated until it keeps up, so magic states are taken to be available on demand~\cite{molavi2025dascot, zhou2026topols, silva2024lssp, watkins2024high}.

Both assumptions can be improved. Cultivation allows a $T$ state to be prepared inside any idle patch. Supply thus scales with the ancilla \emph{area} of the layout rather than its perimeter, raising $T$ parallelism to $\mathcal{O}(N)$~\cite{hofmeyr2025puremagic, hirano2025locality}, and the compiler's question becomes which idle patches to cultivate on. Additionally, the true magic supply rate is a random variable due to post-selections, with a long right tail whose realization is known only at runtime (Section~\ref{sec:msc-model}). Budgeting for the mean stalls on the tail; budgeting for the tail wastes the volume reserved.

Recent schedulers start to treat cultivation latency as a first-class constraint~\cite{hofmeyr2025puremagic, hirano2025locality}, but they place instructions on a fixed 2D layout with a given qubit mapping. The flexibility of cultivating everywhere is thus constrained by the same fixed 2D routing algorithm, and never fully unleashed by co-optimizing with the 3D embedding.

\section{Methods}
Our FlowRouter compiler consists of two parts. The first part routes the Clifford gates at compile time, and the second part schedules the $T$ gates at runtime. Together, they produce a single pipe diagram for both Clifford and $T$ gates reflecting the quantum processor's runtime operation.

\subsection{Compile Time Clifford Routing}
\begin{figure}[t]
    \centering
    \includegraphics[width=0.98\columnwidth]{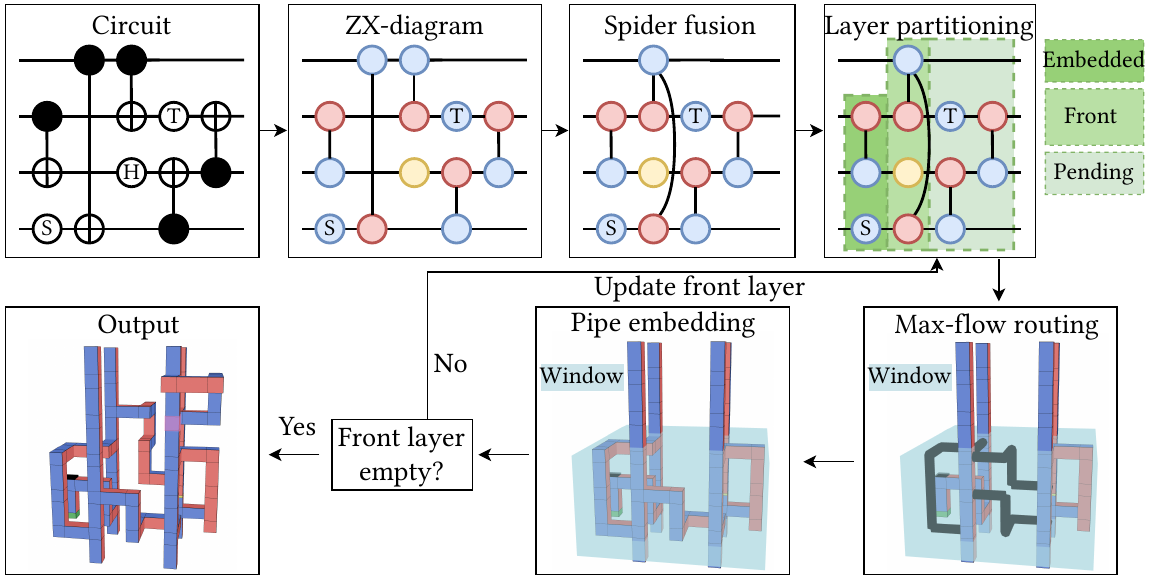}
    \caption{Overview of our max-flow-based Clifford routing. }
    \label{fig:partA}
\end{figure}
We first present our algorithm for routing the Clifford gates into the 3D pipe diagram, and then the algorithm that supplies it with a good initial mapping.
\subsubsection{Layered Max-Flow Routing}
Given an input circuit composed of a standard Clifford+$T$ gate set $\{H, S, \mathrm{CX}, T\}$, the objective is to route the Clifford gates into a 3D pipe diagram of small spacetime volume. This stage runs at compile time. Our method is outlined in Figure~\ref{fig:partA} and detailed below.

We first translate the circuit into its ZX-diagram representation. On this diagram we apply the \emph{spider fusion rule}, merging adjacent spiders of the same type into a single spider. This fusion process increases gate execution parallelism and reduces the spacetime cost. We intentionally restrict our ZX-calculus optimizations to spider fusion to minimize compilation time; applying more sophisticated ZX rewrite rules could further reduce spacetime costs but at a substantially higher computational cost.
 
Second, we partition the ZX diagram vertically into layers, so that the ZX operations within a layer can be executed in parallel. Note that two gates fused into a single spider belong to the same layer. At any point during compilation, each layer is in one of three states: \emph{embedded}, already placed in the 3D pipe diagram and therefore routed; \emph{front}, not yet embedded but with all predecessor layers embedded; and \emph{pending}, still awaiting its predecessors. We process layers front to back, embedding the front layer and then advancing the front.

\begin{figure}[t]
    \centering
    \includegraphics[width=0.98\columnwidth]{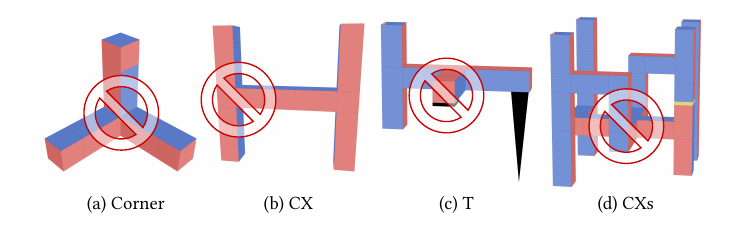}
    \caption{Four violations of inviolable constraints.}
    \label{fig:constraint}
\end{figure}

During embedding, the routing must satisfy several structural constraints to produce a valid pipe diagram. Figure~\ref{fig:constraint} shows example violations. 

\begin{itemize}
    \item No 3d corners may exist.
    \item A CX is composed of a red junction on the target qubit and a blue junction on control (with no Hadamards in between).
    \item Any conditional measurements must occur after (above) the variables they are conditioned on.
    \item Two pipe routes cannot pass through the same location at the same time, so distinct routes must not share any point in spacetime.
   
\end{itemize}

We formulate the embedding as a max-flow problem and solve it with repeated applications of Dinic's algorithm~\cite{dinic1970algorithm}. Each max-flow subroutine is run with a window, where every part of the routes generated are at or above the window's lower bound and at or below the window's upper bound, vertically. Not every path Dinic's finds corresponds to a legal CX gate, so we alternate between running Dinic's to generate new candidate paths and editing existing paths to legalize them. We repeat this embedding across multiple windows, running the \verb|fill| subroutine below in each. Every qubit touched by a front-layer gate is colored red or blue by that gate's spider. Within the window, a red cube above all previously placed blue (and yellow Hadamard) spiders becomes a source; a blue cube  above all previously placed red (and yellow) spiders becomes a sink. Connections leave sources only through red faces and enter sinks only through blue faces.

\verb|fill|: each trial runs Dinic's on the current residual network to find new augmenting paths, then checks each against the CX legality criteria. Legal paths are kept as-is; illegal ones go through repair attempts that insert vertical detours in the middle of the path's xy-junctions (adjacent pairs of pipes in the path such that one points in the $\pm$x-direction and the other in the $\pm$y-direction). We commit paths to a temporary copy of the pipe diagram in order of increasing endpoint height, blocking committed cubes from future commits, and stop after three iterations.

At the beginning of each \verb|fill| call, we initialize a window $w_1$, 3 cubes tall. \verb|fill|'s purpose is to extend the pipe diagram within $w_1$ to include as many CXs still present in the front layer as possible, in order to produce a baseline routing called $r_1$. This process involves discarding max-flow-generated paths that cannot represent CX gates due to some violation as in Figure~\ref{fig:constraint}. If the call to \verb|fill| with $w_1$ places nothing, we retry with a smaller search space: first halving the pool of eligible endpoint qubits and rerunning \verb|fill| repeatedly, down to a floor of two qubit columns; if that still fails, we shift $w_1$ forward one block in time, reset the column pool to all qubits, and repeat. This shrink-and-advance loop is guaranteed to eventually place at least one operation, so $r_1$ is always nonempty.

Starting from $r_1$, we search for a more compact routing by extending the window downward. For each of $r_1$'s source/sink endpoint columns, we compute the deepest block where it could still legally reopen for a new connection; the shallowest such depth across all endpoint columns sets a floor the window cannot pass. Each downwards step halves the remaining gap, lowering the window bottom to that depth to form a candidate $w_i$ and rerunning \verb|fill| on another temporary copy of the pipe diagram. We accept a trial if it places at least half as many CXs as $r_1$, comitting the accepted trial with the deepest floor $r_2$.

Every non-CX gate is placed at the lowest available cube in its qubit's pipe column once it reaches the front layer. Two exceptions add requirements: a $T$ gate reserves an adjacent cube for $T$-state injection (Section~\ref{sec:Tcultivation}), and an $S$ gate runs a breadth-first search outward to locate both the $Y$-measurement site and the classical measurement site above it. Once the current front layer's routing is fixed, we move its ZX operations to the embedded layer, update the front and pending layers, and repeat until every ZX operation is embedded.

\subsubsection{Initial Mapping}

\begin{figure}[t]
    \centering
    \includegraphics[width=0.98\columnwidth]{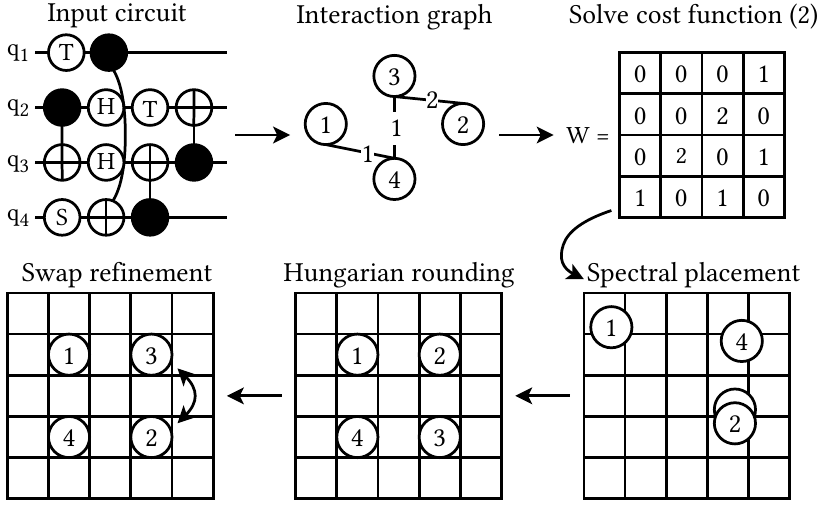}
    \caption{Initial mapping demonstration}
    \label{fig:initial-mapping}
\end{figure}
The quality of the routing solution depends strongly on the initial mapping, which assigns each logical qubit in the circuit to a distinct data patch in the layout. A well-chosen mapping can substantially reduce the spacetime volume of the resulting pipe diagram. Note that unlike qubit mapping for NISQ architectures~\cite{li2019tackling, niu2020hardware,zou2024lightsabre}, this assignment is static: it is determined before execution and is never modified through SWAP operations. We propose the initial-mapping algorithm illustrated in Figure~\ref{fig:initial-mapping}, which proceeds as follows.

Let $G=(V,E,w)$ be the interaction graph of the circuit, where $V$ is the set of logical qubits, $E$ the pairs acted on by at least one two-qubit gate, and $w_{ij}$ the number of such gates on pair $(i,j)$. We fix the sparse layout as shown in Figure~\ref{fig:initial-mapping} (the numbered patches represent data qubits and the remaining patches serve as ancilla qubits for routing) and write $\mathcal{P}=\{(x,y)\in\mathbb{Z}^{2} : x,y \text{ odd},\;
1\le x,y\le 2m-1\}$ for the coordinates of its data patches, where $m$ is chosen so that $|\mathcal{P}|\ge|V|$. An initial mapping is an injection $\pi: V \to \mathcal{P}$, and we choose it to minimize
\begin{equation}
\label{eq:initial-mapping}
\Phi(\pi) \;=\; \sum_{(i,j)\in E} w_{ij}\,\bigl\lVert \pi(i)-\pi(j) \bigr\rVert_1 .
\end{equation}
Shorter distances reduce the routing cost of two-qubit gates, so this 2D objective serves as a proxy for the 3D spacetime volume. Minimizing $\Phi$ is NP-hard, so we solve it with the heuristic described below.

We relax \eqref{eq:initial-mapping} in two ways: qubits may sit at arbitrary real coordinates rather than on the odd-integer patch locations, and the $\ell_1$ Manhattan distance is replaced by the squared $\ell_2$ Euclidean distance. Collecting the coordinates in $x,y\in\mathbb{R}^{|V|}$ and writing $L=D-W$ for the graph Laplacian of $G$ ($W_{ij}=w_{ij}$, $D_{ii}=\sum_j w_{ij}$), the relaxed objective separates as
\begin{equation}
\label{eq:relaxed}
\sum_{(i,j)\in E} w_{ij}\bigl[(x_i-x_j)^2+(y_i-y_j)^2\bigr]
  \;=\; x^{T}Lx + y^{T}Ly ,
\end{equation}
with orthonormal $x,y$. Excluding the trivial solutions that place all qubits at a single point or on a single line, the Rayleigh--Ritz principle shows that \eqref{eq:relaxed} is minimized by the eigenvectors of the two smallest nonzero eigenvalues of $L$~\cite{hall1970r}. We take these as the $x$- and $y$-coordinates of the qubits, rescaled so the points fill the layout region. The entries are arbitrary real numbers, so the resulting coordinates do not generally coincide with patch locations.

To make the coordinates legal patch locations, i.e. odd-integer coordinates in $\mathcal{P}$, we assign each qubit to a distinct patch as close as possible to its relaxed position. Writing $\pi(i)=(\pi_x(i),\pi_y(i))$, we solve
\begin{equation}
\label{eq:assignment}
\min_{\pi}\; \sum_{i\in V} \bigl|x_i-\pi_x(i)\bigr| + \bigl|y_i-\pi_y(i)\bigr|
\end{equation}
over injections $\pi:V\to\mathcal{P}$. This is a linear assignment problem, which we solve using the Hungarian algorithm in $O(|V|^{3})$ time.

To further improve the initial mapping, we perform a local search over pairwise swaps. For each of the $O(n^2)$ pairs of data qubits, where $n=|V|$, we exchange the patches assigned to them and evaluate \eqref{eq:initial-mapping} for the resulting mapping. We apply the swap attaining the lowest $\Phi$ and repeat until no swap improves it, taking the result as our initial mapping.

\subsection{Runtime $T$ gate Scheduling}
\label{sec:Tcultivation}
We first describe how we model the stochastic magic state cultivation process,
and then the delay mechanisms that absorb the cultivation latency into the
Clifford pipe diagram to schedule each $T$ gate at runtime.

\subsubsection{$T$ Cultivation Modeling}
\label{sec:msc-model}
We use MSC (Section~\ref{sec:background}) to prepare high-fidelity $T$ states, and assume cultivation runs in every idle logical patch of the surface code. When a route occupies a patch, the in-progress state is discarded and cultivation restarts from injection once the patch is idle again.

During injection and cultivation only a small fraction of a patch's qubits are occupied. We therefore adopt \emph{in-patch multiplexing}~\cite{kim2026reducing}, in which four cultivation attempts run in parallel in corner-localized regions of a single logical patch. If one or more attempts survive the injection and cultivation stages, one is forwarded to a single shared escape stage and the remainder are dropped. Since only one of the four must survive, this substantially reduces the expected time to produce a $T$ state.

We simulate MSC under the SI1000 noise model~\cite{gidney2021fault}, which more closely matches current superconducting hardware, and we allocate $10\,\mu$s ($10$ cycles at a $1\,\mu$s cycle time) for the complementary gap computation following~\cite{gidney2024cultivation}. Following PureMagic~\cite{hofmeyr2025puremagic}, we run Markov chain Monte Carlo simulations using the per-stage survival probabilities of magic state cultivation reported in Figure~15 of Gidney et al.~\cite{gidney2024cultivation}, mathematically rescaled to account for in-patch multiplexing across the injection and cultivation stages. We compute the expected cultivation time of the resulting process, fit an exponential distribution to it, and draw cultivation completion times from that distribution.

The protocol provides three tunable parameters that trade spacetime volume against the logical error rate (LER) of the resulting $T$ state: the fault distance $d_1$ reached at the end of cultivation, the surface code distance $d_2$ reached after escape, and the complementary gap threshold $G_T$. We sweep $d_1\in\{3,5\}$, $d_2\in\{11,13,15\}$, and $G_T\in\{0,5,\dots,100\}$, selecting $d_1$ and $G_T$ so that the cultivated state's post-escape LER approximately matches that of a distance-$d_2$ surface code patch, ensuring cultivation is not the dominant error source in the computation. For $d_2\in\{11,13\}$, $d_1=3$ with a suitable $G_T$ already satisfies our target error rate. We therefore prefer $d_1=3$ to $d_1=5$, since the shorter cultivation protocol involves fewer postselected checks and consequently has a lower expected preparation time. For $d=15$, increasing $d_1$ to 5 is required to bring the LER down to surface-code levels. The resulting expected cultivation times are given in Table~\ref{tab:cultivation}. We report the cultivation runtime in code cycles, $E(C_t)$, the expected number of syndrome-extraction rounds, and in logical cycles, $\mathrm{Log}_{\mathrm{cycle}} = E(C_t)/d_2$, the same quantity expressed in units of the cube height.

\begin{table}[t]
  \centering
  \caption{Expected cultivation runtimes. $E(C_t)$ is in code cycles;
  $\text{Log}_\text{cycle}$ converts it to logical cycles of $d_2$ code cycles each.}
  \begin{tabular}{l|c|c|c}
    \hline
    \textbf{$d_1$} & \textbf{$3$} & \textbf{$3$} & \textbf{$5$} \\ \hline
    \textbf{$d_2$} & \textbf{$11$} & \textbf{$13$} & \textbf{$15$} \\ \hline
    \textbf{$E(C_t)$} & 23.2     & 26.4          & 82.5     \\ \hline
    $\text{Log}_\text{cycle}$ & 2.1     & 2.0          & 5.5     \\ \hline

  \end{tabular}

  \label{tab:cultivation}
\end{table}

\subsubsection{Delay Mechanisms}
We schedule $T$ gates at runtime since MSC process is stochastic. Since MSC typically takes longer than Clifford gates, when execution reaches a $T$ gate, the $T$ state may not yet be ready or, if one is available, there may be no feasible route connecting it to the gate. In either case, we introduce several delay mechanisms that defer the execution of the $T$ gate. The process is as follows.

We begin at the bottom of the pipe diagram produced for the Clifford gates, which is the temporal starting point, and start an MSC process in every empty region. As cultivation proceeds, we continuously check whether each MSC process has finished. To guarantee that a $T$ state can always reach its $T$ gate, we reserve one ``backup $T$ injection cube" next to each $T$ gate. With this reservation, when execution reaches a $T$ gate, one of two cases occurs:

\textbf{Case 1}: at least one $T$ state is ready a cycle before the $T$ gate is scheduled to be executed. We route from the $T$ gate to the nearest ready $T$ state using breadth-first search, if possible.

\textbf{Case 2}: we reach the $T$ gate but no $T$ state is ready. We apply a delay mechanism to wait until a $T$ state becomes ready, then route it to the $T$ gate.

\begin{figure}[t]
    \centering
    \includegraphics[width=0.98\columnwidth]{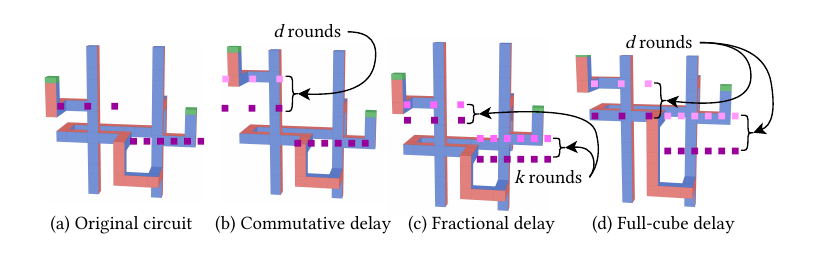}
   \caption{Three delay mechanisms applied to the same reference circuit. 
Dense purple dotted lines mark the first $T$ gate, sparse purple dotted lines
the second, and light purple dotted lines the updated positions after a delay.}
    \label{fig:delay}
\end{figure}

We propose three delay mechanisms to defer the execution of a $T$ gate, illustrated in Figure~\ref{fig:delay} against a single reference circuit with two $T$ gates and two CXs, whose undelayed pipe diagram is shown in Figure~\ref{fig:delay}(a). When a delay is required we try the mechanisms in order of increasing cost, commutative, then fractional, then full-cube, so that we add as little spacetime volume as possible. The delay is applied incrementally, and after each increment we re-check whether a $T$ state is ready and whether a route from the $T$ gate to that state is available, stopping as soon as both hold.

\textbf{Commutative delay.} We move the $T$ gate later in time to open up space for routing, which does not increase the spacetime volume of the Clifford gates. This is valid only when the $T$ gate commutes with the following operations (or the following operation is a Hadamard that itself can be commuted upwards); in our gate set, this holds when the following operation is an idle cube, an $S$ gate, or the control cube of a CX (or a Hadamard with empty space above). As shown in Figure~\ref{fig:delay}(b), we delay the $T$ gates by $d$ rounds by commuting it with the idle cubes.

\textbf{Fractional delay.} Let $t$ be the current program time and $d$ the code distance, and suppose a $T$ gate is scheduled at $t$. We execute the remaining gates of that layer and defer the $T$ gate by one syndrome-extraction round; every circuit element, cube or pipe, that begins at or after $t+d$ is deferred by the same round, so the diagram stays consistent. We repeat this one round at a time, re-checking after each round whether a $T$ state is ready amongst the explored routes, for at most $k \le d-1$ rounds. A fractional delay opens only a fraction of a cube, so it does not create space for a new routing cube; what it buys is time for cultivation to finish. This process is shown in Figure~\ref{fig:delay}(c). If the $T$ state still cannot be routed after $d-1$ rounds, we escalate to the full-cube delay.

\textbf{Full-cube delay.} If $d-1$ rounds of fractional delay do not release the deferred $T$ gate, we escalate to a full-cube delay. Again let $t$ be the current program time, with the $T$ gate scheduled at $t$. We execute the remaining gates of that layer, and defer the $T$ gate together with every element beginning at or after $t+d$ by one full cube. Because this inserts complete cubes rather than a fraction of one, it opens new routing space as well as giving more time for a $T$ state to become ready. This process is shown in Figure~\ref{fig:delay}(d).

Once a delay mechanism has made a $T$ state ready and reachable, we route from the $T$ gate to that state by breadth-first search, and run a second breadth-first search to place the machinery for the conditional $S$ correction that $T$ state injection requires. The entire process is simulated classically on the CPU and completes within the $15\,\mu$s budget at $d = 15$ (Section~\ref{sec:latency}). A successful $T$ gate is shown in Figure~\ref{fig:gates}, where the black component represents MSC. This whole process is applied to every $T$ gate in the circuit.

\subsection{Correctness of the Methods}

To check that the generated pipe diagrams indeed implement the input circuit, we evaluate correctness in two steps: first for the Clifford part of the circuit, then for the $T$ gate part.

For the Clifford part we use \emph{correlation surfaces} (Section~\ref{sec:background}): we generate the pipe diagram with our compiler and pass it to TQEC's \verb|find_correlation_surfaces| function. Each surface $\sigma$ returned by the function carries a Pauli string $P^{\mathrm{in}}\sigma$ supported on the diagram's input ports and a Pauli string $P^{\mathrm{out}}\sigma$ supported on its output ports, expressing that the logical observable $P^{\mathrm{in}}\sigma$ at the input is correlated with $P^{\mathrm{out}}\sigma$ at the output. \verb|find_correlation_surfaces| represents each Pauli operator as a two-bit binary string ($I\leftrightarrow00,X\leftrightarrow 01,Z\leftrightarrow10,Y\leftrightarrow11$), and returns a generating set of all correlation surfaces under the bitwise XOR ($\bigoplus$) operation (represented as a generating set of the set of concatenated input+output strings).

We independently build a second such generating set directly from the circuit. For each qubit $i \in {0,\dots,N_q-1}$ ($N_q$ the qubit count) and each $J \in \{X,Z\}$, let $P_{J,i}$ denote the Pauli string with operator $J$ at index $i$ and $I$ elsewhere. Since Qiskit's \verb|evolve| implements the Heisenberg-picture map $C^\dagger P C$ for a Clifford $C$, we seed $P_{J,i}$ at the diagram's output port $i$ and take the evolved operator as the corresponding input string. We then recover a corresponding input Pauli string for every one of our defined output Pauli strings. Concatenating each seed/evolved pair yields a second generating set for the same input+output space.

Both generating sets are representable as $2N_q \times 4N_q$ binary matrices (there being $2N_q$ strings total, one per choice of $X$ or $Z$ at each of the $N_q$ qubits, each string being $4N_q$ in length from the concatenation of an input block and an output block, each consisting of $N_q$ 2-bit binary substrings), which we reduce to Row-Reduced Echelon Form (RREF) via Gauss–Jordan elimination and compare for equality. This verifies the Clifford pipe diagram up to sign: our two-bit encoding does not track the sign of each Pauli string, so the check certifies structural correctness but not the overall Pauli frame. This is nonetheless a complete notion of correctness, since any sign discrepancy is a global Pauli byproduct that can always be corrected by conjugating the implementation with a fixed Pauli string, and so does not change the logical circuit realized.

For the verification of the T-gates, correlation surfaces do not apply, since they currently certify Clifford propagation only. We therefore assert each $T$ gate by checking that the  prepared $T$ state is injected at a $Z$-type (blue) junction on the data qubit on which that $T$ gate acts.

\section{Evaluation}
\label{sec:eva}
\textbf{Benchmark selection.} We source benchmarks from MQT-Bench~\cite{quetschlich2023mqt}, FTCircuitBench~\cite{harkness2026ftcircuitbench}, and TopoLS, covering a broad variety of quantum algorithms with different sizes. We select for the set of circuits that complete compilation under our time cutoff of 4 hours. Multiple instantiations of a circuit family reflect the circuit family's high diversity in compilation strategies. We compile these circuits (from MQT-Bench and FTCircuitBench) into the Clifford+$T$ gate set using Gridsynth~\cite{rossselinger2016} with an approximation precision of $\epsilon = 10^{-3}$, which is sufficient for the logical error rates of $10^{-6}$ to $10^{-7}$ expected of early fault-tolerant systems~\cite{hao2025trasyn}, and we further optimize them with tzap~\cite{albarghouthi2026linear} to reduce the $T$ gate count. The circuits from TopoLS are already in the same gate set and optimized. Detailed information on the benchmark circuits is given in Table~\ref{tab:benchmark-screening}; Table~\ref{tab:motiv-tratio} summarizes their $T$ gate demand by circuit class.

\begin{table}[t]
    \centering
    \small
    \setlength{\tabcolsep}{4pt}
    \renewcommand{\arraystretch}{.94}

    \caption{Benchmarks after common circuit optimization.}
    \label{tab:benchmark-screening}

    \begin{tabular}{|l|r|r|r|}
        \hline
        Benchmark & Gates & CX & $T/T^\dagger$ \\
        \hline

        GHZ(16)\textsuperscript{3}
            & 16 & 15 & 0 \\

        Adder(10)\textsuperscript{2}
            & 113 & 53 & 32 \\

        Bernstein--Vazirani(20)\textsuperscript{1}
            & 10 & 9 & 0 \\

        Bernstein--Vazirani(150)\textsuperscript{1}
            & 75 & 74 & 0 \\

        CDKM adder(40)\textsuperscript{1}
            & 533 & 248 & 152 \\

        Deutsch--Jozsa(20)\textsuperscript{1}
            & 21 & 19 & 0 \\

        Deutsch--Jozsa(100)\textsuperscript{1}
            & 102 & 99 & 0 \\

        Graph state(20)\textsuperscript{1}
            & 40 & 20 & 0 \\

        Graph state(100)\textsuperscript{1}
            & 200 & 100 & 0 \\

        Grover(6)\textsuperscript{3}
            & 1,015 & 245 & 666 \\

        HHL(4)\textsuperscript{2}
            & 1,489 & 18 & 706 \\

        QAOA(16)\textsuperscript{3}
            & 176 & 48 & 80 \\

        QFT(16)\textsuperscript{1}
            & 7,643 & 210 & 3,558 \\

        QFT(4)\textsuperscript{2}
            & 555 & 12 & 265 \\

        QNN(25)\textsuperscript{1}
            & 6,258 & 24 & 3,027 \\

        VBE adder(25)\textsuperscript{1}
            & 286 & 163 & 64 \\

        VQE(16)\textsuperscript{3}
            & 343 & 30 & 204 \\

        VQE (real amplitudes)(16)\textsuperscript{1}
            & 4,187 & 45 & 1,997 \\

        W state(26)\textsuperscript{1}
            & 3,095 & 51 & 1,466 \\

        \hline
        \multicolumn{4}{@{}l@{}}{\rule{0pt}{1.4em}
            \textsuperscript{1} MQT-Bench;
            \textsuperscript{2} FTCircuitBench;
            \textsuperscript{3} TopoLS.} \\
    \end{tabular} \\
\end{table}

\begin{table}[t]
    \centering
    \small
    \setlength{\tabcolsep}{5pt}
    \caption{$T$ gate demand of our benchmark suite after Clifford$+T$ synthesis,
    grouped by circuit class. The last column is the class-wide ratio of
    $T/T^\dagger$ gates to CXs.}
    \label{tab:motiv-tratio}
    \begin{tabular}{lrrrr}
        \toprule
        Circuit class & \# & CX & $T/T^\dagger$ & $T$ per CX \\
        \midrule
        Clifford-only          &  7 &   336 &      0 & 0.0  \\
        Toffoli-based          &  4 &   709 &    914 &  1.3 \\
        Synthesized rotations  &  8 &   438 & 11,303 & 25.8 \\
        \midrule
        Suite                  & 19 & 1,483 & 12,217 &  8.2 \\
        \bottomrule
    \end{tabular}
\end{table}

\textbf{Experiment Platform.} We evaluate the benchmarks using different compilers on an AWS t3.xlarge instance with $4$ vCPUs (approximately equivalent to $2$ physical CPU cores) and 16~GiB of memory. The experiment demonstrating the scheduler's ability to make rapid decisions at runtime is conducted on a 2.3~GHz quad-core Intel Core i7 processor since the scheduling process is very fast.

\textbf{Algorithm Configuration.}
Clifford routing is performed offline at compile time using a Python implementation with a window size of $w_1=3$. To evaluate online $T$ gate scheduling, we implement the quantum-computer simulator in Python and the latency-critical scheduling engine in Rust. During simulated execution, the Rust scheduler dynamically determines $T$ gate routes and decides whether $T$ gates should be delayed based on the current execution state and resource availability.

\textbf{Baselines.}
We benchmark FlowRouter against the following set of state-of-the-art surface code compilers: TopoLS~\cite{zhou2026topols}, PureMagic~\cite{hofmeyr2025puremagic}, and DASCOT~\cite{molavi2025dascot}. For each compiler, we enable its published settings that maximally reduce its outputs' spacetime volumes: we enable Full-Opt for TopoLS, lightweight PBC ($\omega=1$) for PureMagic, and the mapping-and-routing mode with a sparse layout for DASCOT.

\textbf{Metrics.} We use the following metrics: (1) \emph{spacetime volume} (bounding volume): the product of the spatial footprint and logical execution cycle count of the resulting lattice surgery program; (2) \emph{wall-clock compilation time}: CPU time measured from the submission of the input circuit to the generation of the final pipe diagram; (3) \emph{scheduler latency}: the CPU time spent by the online scheduling engine per logical cycle to determine $T$ gate routes and delay decisions; (4) \emph{application infidelity ratio}: calculated by computing the ratio of the two compilers' $1-(1-e)^a$ terms, where $e$ is the LER of a surface code with distance $d$ and $a$ is the active volume (the number of cubes in the pipe diagram).

\subsection{Clifford Routing Under Free Magic}

In this section, we assume that $T$ states are always available from dedicated magic state factories located along the boundary of the computational region. This is the setting under which TopoLS and DASCOT were originally evaluated, and we extend it to FlowRouter and PureMagic so that all four compilers are compared on identical terms.

\begin{table*}[t]
    \centering
    \small
    \setlength{\tabcolsep}{3.5pt}
    \renewcommand{\arraystretch}{1.05}

    \caption{The spacetime volume and wall-clock compilation time for Clifford routing experiment. Ratio is $V_{\mathrm{baseline}}/V_{\mathrm{FlowRouter}}$, where $V$ is spacetime volume; values above 1 favor FlowRouter.}

    \resizebox{0.8\linewidth}{!}{%

    \begin{tabular}{|l|rr|rrr|rr|rrr|}
        \hline
        \multirow{2}{*}{Benchmark}
        & \multicolumn{2}{c|}{FlowRouter}
        & \multicolumn{3}{c|}{TopoLS Full-Opt}
        & \multicolumn{2}{c|}{PureMagic}
        & \multicolumn{3}{c|}{DASCOT} \\
        \cline{2-11}
        & Volume & Time (s)
        & Volume & Ratio ($\times$) & Time (s)
        & Volume & Ratio ($\times$)
        & Volume & Ratio ($\times$) & Time (s) \\
        \hline

        GHZ(16)
            & \textbf{363} & 0.29
            & \textbf{363} & $\times 1$ & 13.5
            & 1,584 & $\times 4.4$
            & 1,815 & $\times 5$ & 0.03 \\

        Adder(10)
            & \textbf{3,663} & 24.81
            & 7,524 & $\times 2.1$ & 412.2
            & 25,191 & $\times 6.9$
            & 9,438 & $\times 2.6$ & 0.13 \\

        HHL(4)
            & \textbf{18,522} & 165.6
            & N/A & N/A & N/A
            & 25,914 & $\times 1.4$
            & 22,980 & $\times 1.2$ & 1.28 \\

        QFT(4)
            & \textbf{5,880} & 20.76
            & 8,477 & $\times 1.4$ & 1,054.9
            & 12,138 & $\times 2.1$
            & 10,730 & $\times 1.8$ & 0.17 \\

        Bernstein--Vazirani(150)
            & 28,971 & 858.8
            & \textbf{26,622} & $\times 0.9$ & 302.4
            & 45,675 & $\times 1.6$
            & 32,630 & $\times 1.1$ & 0.03 \\

        Bernstein--Vazirani(20)
            & 715 & 1.05
            & \textbf{572} & $\times 0.8$ & 13.5
            & 1,170 & $\times 1.6$
            & 1,089 & $\times 1.5$ & 0.003 \\

        CDKM adder(40)
            & \textbf{28,305} & 1,607
            & 84,150 & $\times 3.0$ & 2410.2
            & 250,512 & $\times 8.9$
            & 99,700 & $\times 3.5$ & 300.8 \\

        Deutsch--Jozsa(100)
            & \textbf{26,979} & 952.4
            & N/A & N/A & N/A
            & 41,400 & $\times 1.5$
            & 52,370 & $\times 1.9$ & 0.04 \\

        Deutsch--Jozsa(20)
            & \textbf{1,573} & 4.63
            & N/A & N/A & N/A
            & 2,340 & $\times 1.5$
            & 3,211 & $\times 2.0$ & 0.005 \\

        Graph state(100)
            & 2,645 & 347
            & 76,705 & $\times 29$ & 1,417.5
            & \textbf{414} & $\times 0.2$
            & 4,232 & $\times 1.6$ & 1.08 \\

        Graph state(20)
            & 429 & 1.95
            & N/A & N/A & N/A
            & \textbf{117} & $\times 0.3$
            & 676 & $\times 1.6$ & 0.12 \\

        QFT(16)
            & \textbf{71,632} & 7,048
            & N/A & N/A & N/A
            & 155,430 & $\times 2.2$
            & 118,500 & $\times 1.7$ & 1,881 \\

        QNN(25)
            & 31,265 & 2,086
            & N/A & N/A & N/A
            & 48,048 & $\times 1.5$
            & \textbf{29,910} & $\times 1.0$ & 1,889 \\

        VBE adder(25)
            & \textbf{11,830} & 215.8
            & 31,603 & $\times 2.7$ & 1,283.6
            & 76,284 & $\times 6.4$
            & 29,410 & $\times 2.5$ & 51.97 \\

        VQE (real amplitudes)(16)
            & \textbf{17,787} & 257.8
            & N/A & N/A & N/A
            & 33,462 & $\times 1.9$
            & 19,120 & $\times 1.1$ & 1,209 \\

        W state(26)
            & \textbf{29,445} & 542.0
            & N/A & N/A & N/A
            & 150,072 & $\times 5.1$
            & 178,000 & $\times 6.0$ & 1,838 \\

        Grover(6)
            & \textbf{21,294} & 1,109
            & N/A & N/A & N/A
            & 61,776 & $\times 2.9$
            & 48,200 & $\times 2.3$ & 9.52 \\

        QAOA(16)
            & \textbf{2,541} & 12.9
            & 8,107 & $\times 3.2$ & 610.4
            & 10,296 & $\times 4.1$
            & 5,082 & $\times 2.0$ & 9.86 \\

        VQE(16)
            & \textbf{2,541} & 6.65
            & 6,413 & $\times 2.5$ & 480.2
            & 8,811 & $\times 3.5$
            & 4,235 & $\times 1.7$ & 28.6 \\

        \hline
        \textbf{Geometric mean}
            & &
            & & $\times2.3$ &
            & & $\times2.1$
            & & $\times2.0$ & \\
        \hline
    \end{tabular}}
    \label{tab:free-magic}
\par\noindent\footnotesize
Bold: lowest volume. N/A: no result within the 4-hour per-circuit cap or compilation error;
geometric means use pairs available for both compilers. 
\end{table*}

The results of the Clifford routing under free magic experiment are located in Table~\ref{tab:free-magic}. Several benchmarks have large qubit or $T$ counts and are slow to compile, so we cap compilation at four hours per circuit. TopoLS exceeds this cap on some benchmarks and raises a compilation error on others; both are reported as N/A.

\textbf{Spacetime volume.} FlowRouter produces the smallest volume on 14 of the 19 benchmarks. Volumes span three orders of magnitude across the suite, so we summarize with the geometric rather than the arithmetic mean: FlowRouter reduces volume by 2.3$\times$, 2.1$\times$ and 2.0$\times$ against TopoLS, Pure-\\Magic and DASCOT respectively.

\textbf{Compilation time.} We compare wall-clock compilation time only against TopoLS and DASCOT, since PureMagic schedules at runtime and has no separate compilation stage. FlowRouter is slower than DASCOT, which routes in two dimensions rather than embedding in three and therefore searches a much smaller space. Against TopoLS, which also embeds in three dimensions, FlowRouter is $10.8\times$ faster in geometric mean over the benchmarks TopoLS completes. This number understates the scalability gap, since it excludes the benchmarks TopoLS fails to scale to.

\begin{table}[t]
    \centering
    \small
    \setlength{\tabcolsep}{3.5pt}
    \caption{Mapping-scheme ablation: spacetime volume by initial mapping
    scheme. `Random' and `Greedy' cells report mean $\pm$ sample spread over three seeds; our `Spectral' is deterministic and is run once. Bold marks the lowest mean volume for each benchmark.}
    \label{tab:mapping-ablation}
    \resizebox{\columnwidth}{!}{%
    \begin{tabular}{lrrr}
        \toprule
        Benchmark & Random & Greedy & Spectral \\
        \midrule
        Grover(6)          & 15,703.3 $\pm$ 325.2   & 16,356.7 $\pm$ 271.9 & \textbf{15,505.0} \\
        QAOA(16)           & 2,079.0 $\pm$ 46.8     & 1,863.0 $\pm$ 81.0   & \textbf{1,782.0} \\
        VQE(16)            & 2,052.0 $\pm$ 123.7    & 1,971.0 $\pm$ 46.8   & \textbf{1,944.0} \\
        QNN(25)            & 16,698.0 $\pm$ 209.6   & \textbf{16,577.0} $\pm$ 121.0 & \textbf{16,577.0} \\
        VBE adder(25)      & \textbf{7,381.0} $\pm$ 121.0 & 8,308.7 $\pm$ 709.0 & 7,623.0 \\
        QFT(4) & \textbf{2,928} $\pm$ 244 & 2,938 $\pm$ 249.3 & 3,075 \\
        HHL(4) & 9,550 $\pm$ 0 & \textbf{8,913} $\pm$ 900.4 & 9,550 \\
        GHZ(16) & 495 $\pm$ 55.5 & \textbf{360} $\pm$ 0 & \textbf{360} \\
        Deutsch-Jozsa(20) & 1,129.3 $\pm$ 103.1 & 1,260 $\pm$ 0 & \textbf{1,080} \\
        Graph state(20) & 711 $\pm$ 12.7 & \textbf{594} $\pm$ 0 & \textbf{594} \\
        W state(26) & 14,586 $\pm$ 1,262.9 & 13,900.3 $\pm$ 1,080.2 & \textbf{11,869} \\
        \midrule
        Geometric mean    & 3,605.4 & 3,445.1 & \textbf{3,323.5} \\
        \bottomrule
    \end{tabular}
    }
    \par\noindent\footnotesize
We exclude Bernstein-Vazirani benchmarks due to its many unused qubits.
\end{table}

\subsubsection{Initial Mapping Ablation}
We evaluate our spectral initial mapping against two alternatives on a subset of our circuit suite. Define $w_{ij}$ to be the weight of an edge between qubits $i$ and $j$, or the number of CX gates between those qubits in the circuit. The first alternative is a uniform random assignment of logical qubits to legal data patches. The second is a greedy construction: it places the highest-degree qubit ($q_i$ with $\sum_j w_{ij}$ maximal) of the interaction graph at a central patch, then repeatedly pops the heaviest remaining edge and, whenever that edge joins a placed qubit to an unplaced one, assigns the unplaced qubit to the free patch minimizing Euclidean distance to its partner. We benchmark one circuit representative for each circuit class under a stricter 5 minute timeout (due to the repeated trials) (Table~\ref{tab:mapping-ablation}).  We report the strict spacetime volume (the smallest bounding box around active qubits) as a metric, and ignore the effects of $T$ state routing to focus on Clifford density only. Both alternative mappings' metrics are averaged over three seeds, while our deterministic spectral initial mapping is only performed once.

The greedy mapping geometric mean improves on that of the random mapping by $\approx4\%$, while the spectral mapping improves on the random mapping by $\approx8\%$. The spectral mapping generates the best initial mapping on $8/11$ benchmarks, reducing volume by up to $19\%$ on W state(26) against the random mapping; on the remaining three, the performance gap is due to limitations of the objective: $\Phi$ is a two-dimensional proxy that rewards short wires (see \eqref{eq:initial-mapping}), whereas volume also depends on how much free space the layout leaves the router, and a placement that minimizes total wire length can crowd the grid enough to force the max-flow stage to detour.

\subsection{Cultivation-aware Runtime Compilation}

In this section we run the full pipeline, compile time Clifford routing followed by runtime $T$ gate scheduling, and charge magic states both the spacetime volume they occupy and their stochastic cultivation latency. TopoLS and DASCOT do not model magic state production, so the comparison is between FlowRouter and PureMagic alone. We also omit the Clifford-only circuits, which contain no $T$ gates and are unaffected by the cultivation model.

\begin{table*}[t]
    \centering
    \small
    \setlength{\tabcolsep}{4pt}
    \caption{Spacetime volume and infidelity for FlowRouter vs.\ PureMagic across code distances
    $d = 11, 13, 15$. Ratio is $V_{\mathrm{PureMagic}}/V_{\mathrm{FlowRouter}}$ and values above 1 favor FlowRouter; Infidelity is $\mathrm{Infd}_{\mathrm{FlowRouter}}/\mathrm{Infd}_{\mathrm{PureMagic}}$ and values below 1 favor FlowRouter.}
    \label{tab:cultivation-res}
    \resizebox{\textwidth}{!}{%
    \begin{tabular}{lrrrrrrrrrrrrr}
        \toprule
        \multirow{2}{*}{Benchmark}
            & \multicolumn{3}{c}{$d=11$} &
            & \multicolumn{3}{c}{$d=13$} &
            & \multicolumn{3}{c}{$d=15$} & \\
        \cmidrule(lr){2-5} \cmidrule(lr){6-9} \cmidrule(lr){10-13}
            & FlowRouter & PureMagic & Ratio ($\times$) & Infd.
            & FlowRouter & PureMagic & Ratio ($\times$) & Infd.
            & FlowRouter & PureMagic & Ratio ($\times$) & Infd. \\
        \midrule
        HHL(4)
            & \textbf{10,750} & 12,740 & $\times 1.2$ & 2.1
            & \textbf{10,420} & 12,780 & $\times 1.2$ & 2.1
            & 13,930 & \textbf{13,260} & $\times 1.0$ & 2.5 \\
        Adder(10)
            & \textbf{2,778} & 15,440 & $\times 5.6$ & 0.4
            & \textbf{2,651} & 15,440 & $\times 5.8$ & 0.3
            & \textbf{3,709} & 15,440 & $\times 4.2$ & 0.5 \\
        QFT(4)
            & \textbf{3,818} & 5,920 & $\times 1.6$ & 1.5
            & \textbf{3,804} & 5,940 & $\times 1.6$ & 1.5
            & \textbf{5,770} & 6,040 & $\times 1.0$ & 2.2 \\
        QAOA(16)
            & \textbf{2,032} & 6,804 & $\times 3.3$ & 0.5
            & \textbf{1,950} & 6,804 & $\times 3.5$ & 0.4
            & \textbf{2,149} & 6,804 & $\times 3.2$ & 0.5 \\
        VQE(16)
            & \textbf{2,975} & 5,859 & $\times 2.0$ & 0.9
            & \textbf{3,016} & 5,859 & $\times 1.9$ & 0.9
            & \textbf{4,552} & 5,670 & $\times 1.2$ & 1.1 \\
        QNN(25)
            & 40,630 & \textbf{32,560} & $\times 0.8$ & 2.2
            & 36,280 & \textbf{31,680} & $\times 0.9$ & 2.3
            & 78,450 & \textbf{41,690} & $\times 0.5$ & 2.5 \\
        Grover(6)
            & \textbf{22,230} & 32,960 & $\times 1.5$ & 1.4
            & \textbf{21,880} & 32,900 & $\times 1.5$ & 1.4
            & \textbf{31,720} & 32,900 & $\times 1.0$ & 1.9 \\
        W state(26)
            & \textbf{31,600} & 106,500 & $\times 3.4$ & 0.5
            & \textbf{31,670} & 106,300 & $\times 3.4$ & 0.5
            & \textbf{57,160} & 108,600 & $\times 1.9$ & 0.8 \\
        VQE$^*$(16)
            & \textbf{21,360} & 22,360 & $\times 1.0$ & 1.7
            & \textbf{19,400} & 22,430 & $\times 1.2$ & 1.8
            & 43,240 & \textbf{26,840} & $\times 0.6$ & 2.1 \\
        VBE adder(25)
            & \textbf{9,086} & 54,450 & $\times 6.0$ & 0.4
            & \textbf{9,233} & 54,450 & $\times 5.9$ & 0.4
            & \textbf{9,978} & 54,450 & $\times 5.5$ & 0.4 \\
        CDKM adder(40)
            & \textbf{22,407} & 189,450 & $\times 8.5$ & 0.2
            & \textbf{21,930} & 189,450 & $\times 8.6$ & 0.2
            & \textbf{32,578} & 189,450 & $\times 5.8$ & 0.3 \\
        QFT(16)
            & \textbf{85,190} & 97,524 & $\times 1.1$ & 1.5
            & \textbf{87,180} & 97,461 & $\times 1.1$ & 1.5
            & 138,900 & \textbf{98,658} & $\times 0.7$ & 1.8 \\
        \midrule
        \textbf{Geometric mean}
            & & & $\times 2.2$ & $0.87$
            & & & $\times 2.3$ & $0.84$
            & & & $\times 1.6$ & $1.08$ \\
        \bottomrule
    \end{tabular}
    }
    \par\smallskip
    {\scriptsize Bold: lower volume in each FlowRouter/PureMagic pair. $^*$real amplitudes. }
\end{table*}

\subsubsection{Spacetime Volume Across Code Distances}
We compare FlowRouter with PureMagic under the cultivation statistics of Table~\ref{tab:cultivation}. Results are reported in Table~\ref{tab:cultivation-res}. FlowRouter produces the smaller volume on $11$ of the $12$ benchmarks at $d = 11$ and $d = 13$ and on $8$ of $12$ at $d = 15$, reducing spacetime volume in geometric mean by 2.2$\times$, 2.3$\times$ and 1.6$\times$ respectively. Infidelity follows the same pattern: FlowRouter reduces it by $12\%$ and $16\%$ at $d = 11$ and $d = 13$ (resp.), but increases it by $8\%$ at $d = 15$.

The results at $d = 11$ and $d = 13$ are close because the expected cultivation time is nearly the same, $2.1$ and $2.0$ logical cycles (resp.). At $d = 15$ it rises to $5.5$ cycles, roughly $2.7\times$ longer, and the improvement decreases accordingly. The reason is due to the different delay mechanisms between PureMagic and FlowRouter. PureMagic schedules on a two-dimensional layout in runtime, so deferring one gate leaves the geometry of the others untouched. FlowRouter embeds in three dimensions with more complex delay mechanisms that affect the geometric dependencies between gates, so long cultivations cost more. We note that MSC is an actively developing protocol, and reductions in cultivation time, for example a cheaper complementary gap computation, would widen FlowRouter's advantage.

\subsubsection{Scheduler Latency}
\label{sec:latency}
\begin{figure}[t]
    \centering
    \includegraphics[width=0.98\columnwidth]{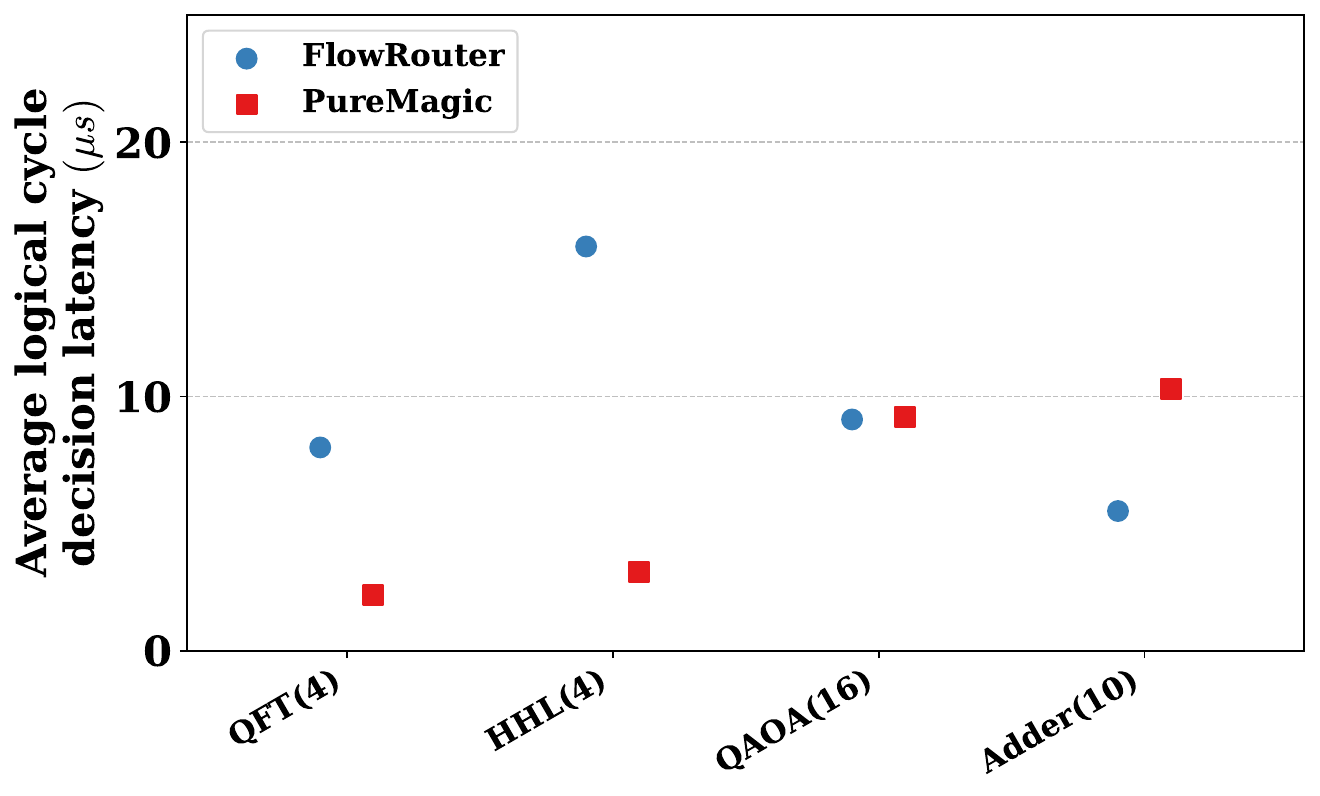}
    \caption{Comparison between PureMagic and FlowRouter execution times per logical cycle.}
    \label{fig:partA-res}
\end{figure}

In order for a processor to make runtime routing decisions, it must be able to act quickly on the current program state. Suppose, for some integer index $i$, $\text{call}_i$ to the routing function returns at time $t=t_0+d$, with $d$ the code distance and $t_0$ the initial time. Define the \textit{program state} at a given time to be a record of all knowable classical or quantum information about the program available at or before that time. If the routing function takes $(n>d)\mu s$ to execute, then $\text{call}_i$ must have been initiated at or before $t-n=t_0+d-n<t_0$. In order for the program to pass a program state into $\text{call}_i$ that indicates which cubes have already been blocked off for routing from times $t_0$ to $t_0+d$ (so that $\text{call}_i$ does not route to logical patches whose $T$ states have been overwritten by $t_0+d$), $\text{call}_{i-1}$ must have finished before $t_0$ (specifically \textit{at or before} $t_0-1$). Repeating this logic, we end up with a buildup of classical processing delays. This implies that the rate at which the instructions are available must be made higher than the rate at which those instructions are dispatched to the quantum processor's control system. Thus, a sufficient condition is that the routing algorithm completes in less than $d\mu s$. In Figure~\ref{fig:partA-res}, we showcase a test of our algorithm's real-time performance using the same average latency per logical cycle metric ~\cite{hofmeyr2025puremagic} employs. We achieve the approximate latency requirement for a $d=15$ surface code on a 2.3GHz Intel quad-core i7. We expect that if implemented on dedicated hardware, our algorithm can achieve significantly lower latency.

\section{Related Work}
\label{sec:related}

\paragraph{Lattice surgery compilers.}
Exact compilers encode an entire embedding as a constraint instance and search for a volume-optimal solution within the modeled space: \texttt{LaSsynth}\\~\cite{tan2024lassynth} over pipe diagrams, and KOVAL-Q~\cite{liao2026kovalq} over a wider family of surface code encodings. Neither scales to whole programs: under a one-day timeout, \texttt{LaSsynth} returns nothing beyond 40 qubits for Bernstein--Vazirani or 8 qubits for random Clifford circuits~\cite{zhou2026topols}. Compilers that do scale route one gate at a time along a shortest available path and differ mainly in how they choose the order~\cite{watkins2024high, leblond2024realistic, molavi2025dascot, hua2021autobraid, silva2024lssp}. The one exception to sequential commitment is EDPC~\cite{beverland2022edpc}, which commits a maximum set of routes jointly by max flow, but only for free-sink magic state delivery; fixed CX endpoints fall back to a greedy $\mathcal{O}(\sqrt{N})$-approximation. TopoLS~\cite{zhou2026topols} is the closest prior work on the routing side: it also expresses the circuit as a ZX diagram and searches 3D pipe diagram embeddings, using Monte Carlo tree search over spider placements under a random spider order per layer, and partitioning large circuits into blocks to scale. FlowRouter shares the layered ZX-to-pipe-diagram formulation, but commits a whole layer in one flow solve instead of searching orders and counts magic state preparation, which TopoLS omits from its reported volume along with the factory footprint. Below the logical level, \texttt{TQEC}~\cite{tqec} and CircLS~\cite{zhang2026circls} lower a pipe diagram or a Pauli product measurement sequence to a physical circuit. At a broader level, end-to-end pipelines co-optimize multiple stages to compile from logical algorithms to physical circuits~\cite{hao2025compilation}. FlowRouter is complementary to these efforts.

\paragraph{Magic state supply and scheduling.}
Because a distillation factory is large and its footprint is fixed~\cite{litinski2019game}, architectures respond structurally: LSQCA tiers patches between a small compute region and dense storage, trading memory access latency for density on the grounds that magic state supply is the binding constraint anyway~\cite{kobori2025lsqca}, while the active volume architecture gives each logical qubit $\mathcal{O}(\log N)$ non-local connections~\cite{litinski2022active}. FLASQ~\cite{huggins2025flasq} estimates the volume an idealized fluid allocation of ancilla patches would need. Cultivation~\cite{gidney2024cultivation} removes the factory altogether by preparing $T$ inside a single surface code patch, in exchange for a completion time known only at runtime. PureMagic~\cite{hofmeyr2025puremagic} exploits this by letting every ancilla patch cultivate whenever it is not routing and canceling cultivation when the patch is needed for a route; locality-aware Pauli-based computation~\cite{hirano2025locality} models preparation latency as a geometric random variable and propagates the resulting delay through the schedule. Both schedule onto a fixed 2D layout with per-instruction path search and take the qubit mapping as given, so readiness constrains a schedule built after routing has been fixed. FlowRouter instead leaves cultivation room inside the 3D embedding by construction and resolves the residual stochasticity by deforming that diagram locally at runtime. SPARO~\cite{kan2025sparo0} balances the allocation of PBC compute, routing, and distillation per workload at compile time. FlowRouter treats Clifford routing and cultivation as a single compilation problem and allocates individual idle patches inside a 3D embedding with runtime resolution.

\section{Conclusion}
In this work we introduced \emph{FlowRouter}, the first topological lattice surgery compiler to place Clifford routing and stochastic magic state cultivation inside one 3D embedding. Across our benchmarks, FlowRouter substantially reduces spacetime volume relative to prior compilers in both the free-magic and cultivation-aware settings. The delay mechanisms FlowRouter uses to absorb cultivation latency also point to a natural next step: co-designing routing and cultivation. A compiler could shape the Clifford pipe diagram so that idle volume appears where and when cultivation needs it, reducing the delay a  $T$ gate incurs. Circuit synthesis offers a second lever, since moving  $T$ gates in the circuit changes when magic states are demanded and therefore how often a delay is required.

\section*{Acknowledgment}
We thank Sumeet Shirgure for the suggestions on the manuscript.
This material is based upon work supported by the U.S. Department of Energy, Office of Science, Office of Advanced Scientific Computing Research under Award Number No.DE-SC0026525.

\bibliographystyle{ACM-Reference-Format}
\bibliography{main}

\end{document}